 \documentclass[final,5p,times,twocolumn]{elsarticle}

\usepackage{amssymb}
\usepackage{amsmath} 
\usepackage{amsfonts}
\usepackage{graphicx}
\usepackage{float}
\usepackage{color}
\usepackage{accents}

\journal{J. Magn. Reson.}

\begin{document}

\begin{frontmatter}



\title{Solid effect DNP polarization dynamics in a system of many spins}


\author{Daniel Wi\'sniewski, Alexander Karabanov, Igor Lesanovsky, Walter K\"ockenberger}

\address{Sir Peter Mansfield Magnetic Resonance Centre, School of Physics and Astronomy, University of Nottingham, University Park, Nottingham NG7 2RD, United Kingdom}

\begin{abstract}
We discuss the polarization dynamics during solid effect dynamic nuclear polarization (DNP) in a central spin model that consists of an electron surrounded by many nuclei. To this end we use a recently developed formalism and validate first its performance by comparing its predictions to results obtained by solving the Liouville von Neumann master equation. The use of a Monte Carlo method in our formalism makes it possible to significantly increase the number of spins considered in the model system. We then analyse the dependence of the nuclear bulk polarization on the presence of nuclei in the vicinity of the electron and demonstrate that increasing the minimal distance between nuclei and electrons leads to a rise of the nuclear bulk polarization. These observations have implications for the design of radicals that can lead to impoved values of nuclear spin polarization. Furthermore, we discuss the potential to extend our formalism for more complex spin systems such as cross effect DNP.
\end{abstract}


\end{frontmatter}


\section{Introduction}
Dynamic nuclear polarization (DNP) is a method  for transferring spin polarization from radical electrons to surrounding nuclei to create a highly polarized non-thermal state which subsequently can be used to acquire a strongly enhanced NMR signal. Several applications of DNP have recently highlighted the huge potential that this method offers for increasing the low sensitivity of magnetic resonance imaging (MRI) and spectroscopy experiments \cite{prigri, atskoc, maly}. In particular, applications of dissolution DNP \cite{arde03, golm03} to prepare highly polarized ${}^{13}$C labelled molecules in conjuction with spectroscopic MRI have led to the development of novel experimental protocols for human cancer diagnostics \cite{Day07, Nelson13}. Despite the recent success of the technique, its mechanism is not yet fully understood and significant effort has gone into developing suitable theoretical models to predict the polarization dynamics during DNP. Most model simulations have focused on the numerical analysis of spin systems that consist of only a few spins \cite{Vega1,Vega2} since the dimension of the required state space of the quantum mechanical problem scales exponentially in the number of constituents and the available computer memory limits the calculations to only up to 10 - 14 spins. In addition, thermodynamical models based on the concept of spin temperature have been used to analyze the dependence of the spin polarization on experimental parameters \cite{Cox77, Wenckebach74, Abragam82} as well as several attempts have been made to describe the polarization dynamics by a set of rate equations \cite{Smith08, Banerjee13, Hovav15}.  We have recently developed a formalism that allows simulations of the polarization dynamics in large many-spin systems for solid effect DNP (SE DNP) \cite{Karabanov2015}. The formalism opens up the possibility of analysing polarization transport in systems consisting up to a few thousand coupled spins. In this contribution we demonstrate the insight that such simulations can provide for understanding the dynamics of many coupled spins during DNP. In particular, we show how the bulk polarization level crucially depends on the molecular environment of the unpaired electron. Furthermore, we include a comparison of our formalism with conventional simulation methods based on solving the quantum mechanical master equation in the full Liouville space. In addition, we discuss the prospects to extend the formalism to more complex DNP pathways in systems with interacting electrons and nuclear spins.
\section{Full quantum mechanical description}
SE DNP can be observed in systems where the electrons have little or no $g$-anisotropy and their concentration is relatively low, resulting in weak dipolar coupling between them and narrow ESR spectra with a linewidth less than the Zeeman splitting of the surrounding nuclear spins \cite{Vega1, jeff57, abpro58, schje65, borgab59}. A quantum model, representative for a spin system in which SE DNP can occur, is the central spin model, where an unpaired electron $S$ is surrounded by an ensemble of $M$ nuclei $I_k, \quad k = 1 ,\dots, M$ \cite{stan14, diet15}. The nuclei are coupled by the hyperfine interaction to the electron and form also a dipolar coupled network between themselves. Polarization can be transferred to nuclei directly surrounding the electron (core nuclei) by applying a microwave field off-resonance in respect to the electron Larmor frequency and on resonance with either the zero (ZQ) or double quantum (DQ) transitions  $\omega_S \pm \omega_I$ of electron-nuclei spin pairs \cite{Vega1,jeff57, abpro58, schje65, borgab59}. The polarization is further distributed between the bulk nuclei via spin diffusion mediated by inter-nuclear dipolar interactions in the spin network. The Hamiltonian in the reference frame of the microwaves
\begin{eqnarray}
\hat{H} &=& \Delta \hat{S}_z +  \sum_k \left (\omega_I \hat{I}_{kz} + A_k \hat{S}_z\hat{I}_{kz} + \tfrac{1}{2}B_{k+}\hat{I}_{k+}\hat{S}_z\right. \nonumber\\
&+& \left. \tfrac{1}{2}B_{k-}\hat{I}_{k-}\hat{S}_z \right) +\tfrac{\omega_A}{2} \left(\hat{S}_+ + \hat{S}_-\right) \label{eq_system_hamiltonian}.
\end{eqnarray}
$\Delta$ represents the microwave offset in respect to the electron Larmor frequency, $\omega_I$ is the nuclear Larmor frequency, $A_k$ is the strength of the secular part  and $B_{k\pm}$ is the strength of the pseudo-secular part of the hyperfine interaction between the electron $S$ and the $k$-th nuclei $I_k$, and $\omega_A$ is the strength of the applied microwave field. Relaxation is introduced by a Lindbladian dissipator, which has the general form: $\textbf{D}\left(\hat{L}_k\right)\hat{\rho} = \hat{L}_k\hat{\rho}\hat{L}_k^\dagger - \left\{\hat{\rho},\hat{L}_k^\dagger\hat{L}_k\right\}/2$ with a set of jump operators defined in the basis of $L_k \in \left\{\hat{S}_\pm, \hat{S}_z, \hat{I}_{k\pm}, \hat{I}_{kz} \right\}$, each with an associated effective rate \cite{us1b}. 
The Lindbladian is commonly used in quantum optics and in the description of dissipation in open quantum systems \cite{lind76, Breuer07}. The dissipator describes relaxation as jumps between different states with effective jump rates that are in our case related to the longitudinal relaxation rates $R_1$, and dephasing terms with effective rates that are given by the transverse relaxation rates $R_2$ observed in the spin system. Using this dissipator form it is straight-forward to introduce more complicated relaxation pathways for the system (e.g. spin cross-relaxation). The action of the Lindbladian dissipator is identical to the use of double-commutator relaxation superoperators, which can be illustrated by the following simple rearrangement. If we take a generic operator $\hat{X}$ (which could for example be one of the lowering or raising operators) along with its hermitian conjugate indicated by the dagger symbol, we can write it in a double-commutator form of a relaxation superoperator
\begin{gather*}
\mathbf{\Gamma}\hat{\rho} = -\frac{R_1}{2}\left(\left[\hat{X},\left[\hat{X}^\dagger,\hat{\rho}\right]\right] + \left[\hat{X}^\dagger,\left[\hat{X},\hat{\rho}\right]\right]\right) \\
  \equiv -\frac{R_1}{2}\left( -2\hat{X}\hat{\rho} \hat{X}^\dagger -2\hat{X}^\dagger\hat{\rho} \hat{X} + \hat{X}\hat{X}^\dagger\hat{\rho} + \hat{\rho}\hat{X}\hat{X}^\dagger + \hat{X}^\dagger \hat{X}\hat{\rho} + \hat{\rho}\hat{X}^\dagger\hat{X}\right)\\
  \equiv R_1 \left(\hat{X}\hat{\rho} \hat{X}^\dagger -\left\{\hat{\rho},\hat{X}^\dagger\hat{X}\right\}/2  + \hat{X}^\dagger\hat{\rho} \hat{X} -\left\{\hat{\rho},\hat{X}\hat{X}^\dagger\right\}/2\right)\\
	\equiv R_1 \left( \textbf{D} \left(\hat{X} \right) + \textbf{D}\left (\hat{X}^{\dagger} \right) \right) \hat{\rho}
\end{gather*}
The Lindbladian is therefore analogous to the Redfield relaxation superoperator. No normalisation term is required with the Lindbladian for the system to relax to thermal equilibrium, as the normalisation term may be included in the effective rate using the principle of detailed balance.
In \cite{us1b} we discussed the differences in simulations seen between using a Lindblad style dissipator defined in the Zeeman basis, and the relaxation superoperator form used by Hovav et. al. \cite{Vega1} in the eigenbasis of the stationary Hamiltonian. Furthermore, we describe in \cite{us1b} the possibility of addition of other dissipation parts to the relaxation superoperator.
The operators $S_\pm$, $I_{k\pm}$ are responsible for transitions between eigenstates of the system and introduce longitudinal relaxation in the system. The operators $\hat{S}_z$, $\hat{I}_{kz}$ are mainly responsible for decay of coherences in the system. Written explicitly for SE DNP, the dissipator arising from a standard random fluctuation model for relaxation in the solid state  has the form \cite{us1b}:
\begin{eqnarray}
\bf{\Gamma}\hat{\rho} &=& \frac{R^{(I)}_1}{2} \sum_k^M \textbf{D}\left(\hat{I}_{k\pm}\right)\hat{\rho} \nonumber \\
&+& \frac{R_1^{(S)}}{2}\left(1\pm p_0 \right) \textbf{D}\left(\hat{S}_{\pm}\right) \hat{\rho}\nonumber \\
&+& 2{R_2^{(I)}} \sum_k^M \textbf{D}\left(\hat{I}_{kz}\right) \hat{\rho}\nonumber\\
&+& 2R_2^{(S)} \textbf{D}\left(\hat{S}_{kz}\right) \hat{\rho}\label{eq_system_relaxation2},
\end{eqnarray}
where $p_0$ is the electron thermal equilibrium polarization, $R_1^{(S)}, R_1^{(I)}$ are the rates of longitudinal relaxation to thermal equilibrium of the electron and nuclei respectively, and $R_2^{(S)},R_2^{(I)}$ the corresponding rates of transverse relaxation. The nuclear thermal equilibrium polarization is assumed to be negligible compared to the electron thermal polarization. A bold font is used for superoperators throughout the text.  \\ 
The spin dynamics are simulated by solving the Liouville von Neumann master equation acting in Liouville space
\begin{equation}
\dot{\hat{\rho}} = -i\left[\hat{H},\hat{\rho}\right] + \bf{\Gamma}\hat{\rho}. \label{eq_system_lindbladian}
\end{equation}

\section{Classical dynamics in the Zeeman subspace}
We have recently demonstrated that the dynamics of the nuclear spin polarization during SE DNP can be well approximated by using only the states contained in the Zeeman subspace \cite{Karabanov2015}. This subspace is spanned by the operators $\left\{1, \hat{S}_z, \hat{S}_z\hat{I}_{kz},\hat{S}_z\hat{I}_{kz}\hat{I}_{k'z}\dots  \right\}$. The effective master equation in the Zeeman subspace can be written in the Lindblad form \cite{lind76, Breuer07} using four jump operators and their corresponding effective rates, representing only four different quantum jumps: electron and nuclear single spin flips, flip-flops between the electron and nuclear spins and flip-flops between spins in nuclear spin pairs. The details of the mathematical procedure can be found in \cite{Karabanov2015}. There we also discuss the advantages of many-body spin simulations, particularly when looking at diffusion of polarization into the bulk of a sample. Limitations of the procedure and the conditions when it can be applied are also provided. The effective master equation has the form 
\begin{eqnarray}
\dot{\hat{\rho}}_{_Z} &=& \sum_k^M \Gamma_{k\pm}^{(I)} \textbf{D} \left(\hat{I}_{k\pm} \right) \hat\rho_{_Z} \nonumber\\ 
&+&  \Gamma_{\pm}^{(S)} \textbf{D} \left(\hat{S}_\pm \right)  \hat\rho_{_Z}   \nonumber \\
&+& \sum_k^M\Gamma_{k}^{(IS)}   \textbf{D}\left(\hat{Y}_k \right) \hat\rho_{_Z} \nonumber \\
&+& \sum_{k<j}^M\Gamma_{k,j}^{(II)} \textbf{D} \left(\hat{X}_{kj} \right)\hat{\rho}_{_Z},
\label{eq_Zeeman_master_equation}
\end{eqnarray}

where $\hat{Y} = \hat{I}_{k+}\hat{S}_- + \hat{I}_{k-}\hat{S}_+$ and $\hat{X}_{kj} = \hat{I}_{k+}\hat{I}_{j-} + \hat{I}_{k-}\hat{I}_{j+}.$ The dimension of the density operator in this master equation scales as $2^{N}$, where $N = M + 1$, with  $M$ given by the number of nuclei in the system. This is a significant reduction in comparison to the full Liouville space which has dimension $4^N$. The effective evolution of the spin polarization in the central electron system is completely classical in the sense that the evolution relies on classical state-dependent rate equations describing the evolution of system polarization, and it is only based on single spin flips and two spin flip-flops and does not require the calculation of any quantum coherences. Hence the evolution connects only classical states in the Zeeman basis and does not create any superpostions unlike a general quantum Master equation dynamics. The effective rates $\Gamma_{k\pm}^{(I)},\; \Gamma_{\pm}^{(S)},\; \Gamma_{k}^{(IS)}$  and $\Gamma_{k,j}^{(II)}$ describe the rates by which these single spin flips and two spin flip-flops occur in the quantum system. An analysis of the mathematical form of the effective rates reveals the dependence of the polarization dynamics on the various system parameters such as the hyperfine interaction strengths $A_k$ and $B_k$, the microwave field strength $\omega_A$ and the relaxation time constants.  Moreover, the effective rates can be used to distinguish between bulk and core nuclei.

The effective rates for the single-spin flips are for the  electron and individual nuclear spins, respectively: 
\begin{gather}
\Gamma_\pm^{(S)} = \tfrac{1}{2} \left(1\mp p_0\right)R_1^{(S)} + \frac{\omega_A^2}{2\omega_I^2}R_2^{(S)},\label{eq_electron_effective_relaxation}\\
\Gamma_{k\pm}^{(I)} = \tfrac{1}{2}R_{1k}^{(I)} + \frac{\left|B_k\right|^2}{8\omega_I^2}R_{2k}^{(I)}.\label{eq_nuclear_effective_relaxation}
\end{gather}
Eq. (\ref{eq_electron_effective_relaxation}) represents the rates for up and down jumps of the electron spin. The different weighting of these two processes arises from the requirement that the electron should relax due to  longitudinal relaxation to its thermal equilibrium value $p_0$. The additional term in Eq. (\ref{eq_electron_effective_relaxation}) stems from the microwave irradiation which, due to the induced Rabi oscillations, can also lead to re-orientation of the electron spin. Similarly, Eq. (\ref{eq_nuclear_effective_relaxation}) describes the effective longitudinal relaxation of the nuclei in the Zeeman frame. The perturbation in this case is due to the coupling with the electron. A more detailed analysis of the second order correction terms can be found in \cite{Karabanov2015}. \\
The dissipative processes acting on two spins describe polarization transfer through flip-flop jumps either involving the electron and a nucleus or two nuclei:
\begin{gather}
\Gamma_{k}^{(IS)} = \frac{\omega_A^2\left|B_k\right|^2\left(R_{2k}^{(I)} + R_2^{(S)}\right)}{8\omega_I^2}\left[\left(R_{2k}^{(I)} + R_2^{(S)}\right)^2 + \hat{D}_k^2\right]^{-1} \label{eq_hyperfine_effective_interaction}\\
\Gamma_{k,j}^{(II)} = \frac{d_{kj}^2\left(R_{2k}^{(I)} + R_{2j}^{(I)}\right)}{2}\left[\left(R_{2k}^{(I)} + R_{2j}^{(I)}\right)^2 + \hat{C}_{kj}^2\right]^{-1}, \label{eq_dipolar_effective_interaction}
\end{gather}
where $\hat{D}_k$ and $\hat{C}_{kj}$ are operator valued terms
\begin{gather}
\hat{D}_k =\lambda + \tfrac{\omega_A^2}{2\omega_I} - \tfrac{\left|B_k\right|^2}{8\omega_I}+\sum_{s\neq k}A_s\hat{I}_{sz}\label{eq_effective_rate_D}\\
\hat{C}_{kj} = \left(A_k - A_j\right)\hat{S}_z + \tfrac{1}{8\omega_I}\left(\left|B_k\right|^2 -\left|B_j\right|^2 \right).\label{eq_effective_rate_C}
\end{gather}
Eq. (\ref{eq_hyperfine_effective_interaction}) describes transfer of polarization from the source electron to the surrounding nuclei. This process depends on the microwave field amplitude squared $\omega_A^2$, the pseudo secular hyperfine coupling strength $B_k$ magnitude squared, the nuclear Larmor frequency squared $\omega_I^2$, the rates of transverse relaxation of the electron $R_2^{(S)}$ and coupled nucleus $R_{2k}^{(I)}$, and the operator-valued term in Eq. (\ref{eq_effective_rate_D}). This term depends on the spin state of the nuclear ensemble which can be intuitively understood because the effective rate for polarization transfer from the electron must at steady-state tend towards the rate by which polarization is lost in the nuclear ensemble due to longitudinal relaxation. In cases of a mismatch of the microwave carrier frequency with either the ZQ or DQ transition frequency, the offset parameter $\lambda$ is large and  the polarization transport efficiency is decreased. \\
Eq. (\ref{eq_dipolar_effective_interaction}) is the effective rate responsible for diffusion of polarization away from core nuclei close to the electron to the bulk. Polarization transfer is driven by nuclear flip-flops, the rate magnitude is dependent on the strength of the dipolar coupling squared $d_{kj}^2$. The presence of $R_2$ values implies that the underlying quantum-mechanical process is affected by transverse relaxation. The effective flip-flop rate is also reduced in cases where there is a large difference between the secular coupling parameters $A_k$ of two nuclei. This implies, that in the case of nuclei very close to the electron, their transport of polarization into the bulk of the sample is inefficient.\\
The second part of Eq. (\ref{eq_effective_rate_C}) is much smaller than the first, and hence $\hat{C}_{kj}$ can be approximated as $(A_k - A_j )\hat{S}_z$, the square of which is $\left(A_k - A_j\right)^2/4 $ and therefore Eq. (\ref{eq_dipolar_effective_interaction}) becomes independent of the spin state of the electron. 
\begin{gather*}
\Gamma_{k,j}^{(II)} = \frac{d_{kj}^2}{2\left(R_{2k}^{(I)} + R_{2j}^{(I)}\right) } \left[1 + \frac{1}{4}\left(\frac{A_k - A_j} {R_{2k}^{(I)} + R_{2j}^{(I)}} \right)^2\right]^{-1}.
\end{gather*}
Comparison between the two rates (Eqs. (\ref{eq_hyperfine_effective_interaction}) and  \ref{eq_dipolar_effective_interaction}) for a nucleus determines whether it belongs to the core or to the bulk. In the case that the effective hyperfine rate, Eq. (\ref{eq_hyperfine_effective_interaction}), is greater than the effective nuclear dipolar rate, Eq. (\ref{eq_dipolar_effective_interaction}), the interaction with the electron will be dominating, indicating a core nucleus. The opposite would be true for a bulk nucleus. 

\subsection{Comparison between Zeeman projection and Liouville von Neuman equation}
\begin{figure}[H]
\includegraphics[scale=0.25]{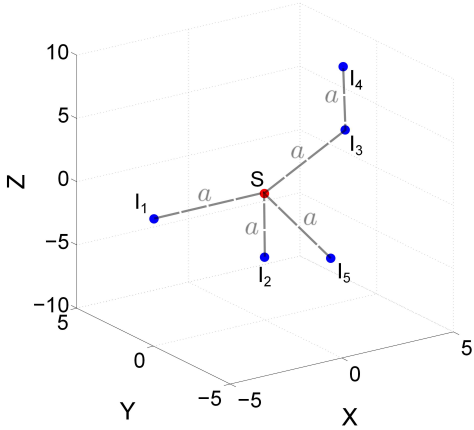}
\caption{Geometry of the model system consisting of one electron ($S$) and five nuclei ($I_k, k = 1, \dots, 5)$. The distance parameter $a$ was varied between 2 \AA \; and 10 \AA \; in a set of simulation to compare the results obtained using the master equation derived from the Zeeman subspace projection with the results of the Liouville von Neuman equation.}\label{fig_test_system_geometry}
\end{figure}

\begin{figure}[H]
\includegraphics[scale=0.2]{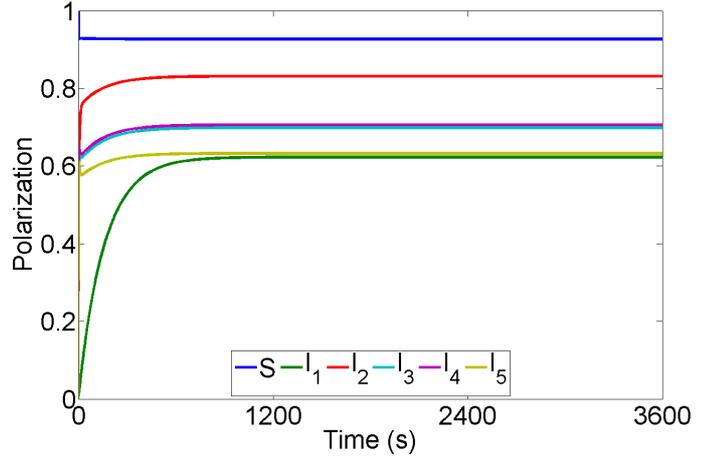}\caption{Polarization curves for spin separation of 2 \AA. Details of the simulation parameters used are provided in the text.}\label{fig_2A_polarization}
\end{figure}

In the following we use a model system to demonstrate the quality of the agreement between simulations based on the Liouville von Neumann equation (Eq. \ref{eq_system_lindbladian}) and an evolution in the full Liouville state space and simulations based on the master equation (Eq. \ref{eq_Zeeman_master_equation}) which was obtained by the projection to the Zeeman subspace. The detailed derivation of the Zeeman-subspace effective master equation in \cite{Karabanov2015} provides the strict, analytical form of conditions which need to be satisfied for the projection to be valid. Here we demonstrate a ficticious system for which we quantitatively verify the parameter regime in which the projection is accurate, with respect to the full master equation. The geometry of the spin system consisting of one electron and five nuclei is shown in Fig. \ref{fig_test_system_geometry}. The system was selected to represent a variety of orientations of nuclei with respect to the electron. Most of the spins are in the $x-z$ plane, as azimuthal changes do not affect the coupling parameters. All the coupling parameters between nuclei as well as those between nuclei and the central electron were obtained from first-principle calculations based on their positions. 
Here $a$ is a variable distance between spins which was varied in simulations between 2 \AA\;  and 10 \AA. The error $\delta p$ was then computed by taking the difference between the polarization values $p_{_\mathrm{LvN}}$ obtained from the Liouville von Neuman master equation (\ref{eq_system_lindbladian}), and the Zeeman-projected master equation (\ref{eq_Zeeman_master_equation}) $p_{_\mathrm{Zeeman}}$,  normalized to the results of the Liouville von Neuman equation, i.e.
\begin{equation}
\delta p = \left(p_{_\mathrm{LvN}} - p_{_\mathrm{Zeeman}} \right) /p_{_\mathrm{LvN}}.\label{eq_polarization_error_definition}
\end{equation}
The coordinates of the nuclei seen in Fig. \ref{fig_test_system_geometry} are listed in Tab. \ref{table1}. For simulations a uniform 5\% randomization in position was used.
\begin{table}[H]
\begin{tabular}{|c|c|c|c|} \hline
\textbf{spin} & $x$ & $y$ & $z$  \\ \hline
$S$ & 0 & 0 & 0  \\ \hline
$I_1$ & $-a$ & 0 & 0 \\ \hline
$I_2$ & 0 & 0 & $-a$ \\ \hline
$I_3$ & $\frac{a}{\sqrt{2}}$ & 0 & $\frac{a}{\sqrt{2}}$ \\ \hline
$I_4$ & $\frac{a}{\sqrt{2}}$ & 0 & $a$$\left(1 + \frac{\sqrt{2}}{2}\right)$ \\ \hline
$I_5$ & 0 & $-a\frac{\sqrt{3}}{2}$ & -$\frac{a}{2}$ \\ \hline 
\end{tabular}
\caption{Coordinates of the electron and the nuclei in the model configuration} \label{table1}
\end{table}
The simulations were carried out using the gyromagnetic constant of the $^1$H nucleus. It was necessary to use  a microwave strength $\omega_A = 500$ kHz in conjunction with the distance parameter value $a = 2$ \AA\; to generate a significant nuclear polarization enhancement. Such microwave strength is difficult to experimentally implement without a cavity. The remaining parameters were $T = 1$ K, $B_z = 3.4$ T, $T_1^{(S)} = 1$ ms, $T_1^{(I)} = 1$ h, $T_2^{(S)} = 200$ ns, $T_2^{(I)} = 0.2$ ms, simulation duration = $1$ h. This parameter set was used for consistency for all simulations with $a$ between 2-5 \AA.  Curves showing the polarization dynamics for the nuclei and the electron are shown in Fig. \ref{fig_2A_polarization}. 
\begin{figure}[H]
\includegraphics[scale=0.2]{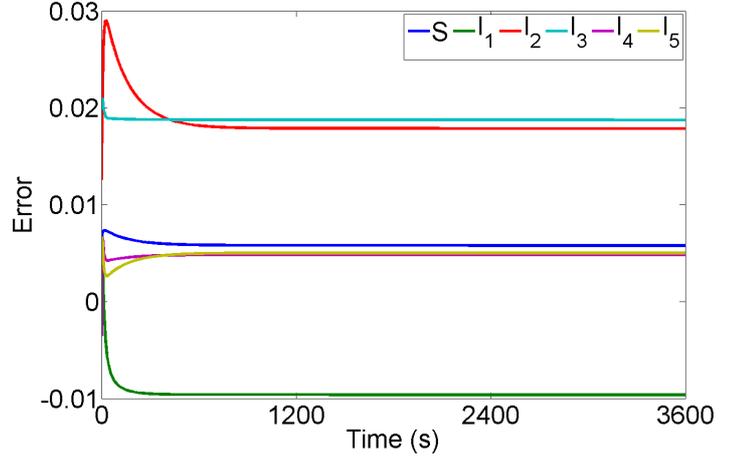}\caption{Error between the calculation based on the Liouville von Neuman equation and the Zeeman master equation for a model spin system with separation parameter value of 2\AA }\label{fig_2A_error}
\end{figure}
The error in polarization for this configuration of the model spin system is shown in Fig. \ref{fig_2A_error}. It does not exceed 3\%. If the separation parameter $a$ is set to 3 or 4 \AA, the error decreases further, e.g. for 4 \AA\; the error never exceeds 1.5\%. For separations of 5-10 \AA\; the microwave power was reduced to an experimentally more realistic strength of $\omega_A = 200$ kHz and the electron transverse relaxation time constant $T_2$ was increased 5-fold to $T_2^{(S)}  = 1\mu$s. All other parameters were kept the same as described previously. The maximum error never exceeded 1\% for any value choice of $a$ between 5 \AA\; and 10 \AA. In all cases of the error analysis, the steady-state error was lower than the maximum error, which usually was found for the data in the initial fast polarization build-up phase. 

In conclusion, generally  a very good agreement is obtained between Eq. (\ref{eq_system_lindbladian}) and Eq. (\ref{eq_Zeeman_master_equation}) for a wide range of parameters that reflect realistic experimental conditions. However, it is worth mentioning that since a set of conditions needs to be fulfilled for the Zeeman projection to provide an acceptable approximation \cite{Karabanov2015}, there are also parameter regimes where the error is likely to be greater than shown here. In particular, the dynamics in model spin systems with very small average distances (below 2\AA) or long transverse relaxation time constants is not well approximated by an adiabatic elimination procedure to project the dynamics onto the Zeeman subspace. In this case coherences are not negligible for the evolution of the spin system and need to be taken into account by including states outside of the Zeeman subspace into the calculations.
 
\subsection{Kinetic Monte Carlo simulations}
\begin{figure}[H]
\includegraphics[scale=0.25]{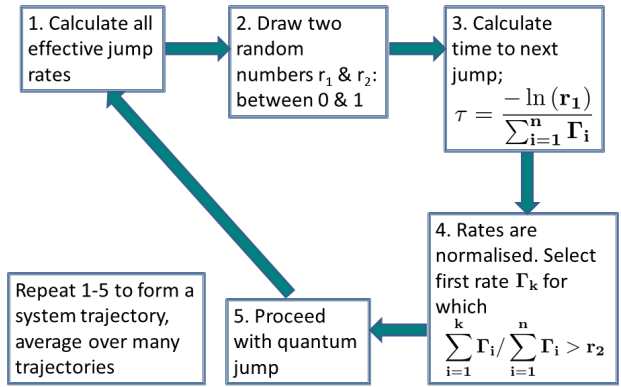}\caption{Flow diagram of the kinetic Monte Carlo algorithm}\label{fig_dmc_schematic}
\end{figure}

The dynamics of spin systems containing a large number $N$ of coupled spins  cannot be calculated by conventional propagation using Eq. (\ref{eq_Zeeman_master_equation}) because the dimensions of the required state space scales with $2^N$. However, because of the classical nature of Eq. (\ref{eq_Zeeman_master_equation}), the dynamics can be well approximated using a kinetic Monte Carlo (KMC) method \cite{Voter07}. It provides a CPU efficient way to find numerical solutions for the evolution of the density operator in Eq. (\ref{eq_Zeeman_master_equation}). Like the static variants, KMC relies on averaging over many trajectories that correspond to possible evolutions of the system. The Zeeman-projected master equation (\ref{eq_Zeeman_master_equation}) is already in a form suitable for this method. The effective rates in Eqs. (\ref{eq_electron_effective_relaxation} - \ref{eq_dipolar_effective_interaction}) along with their corresponding Lindblad operators can be directly used. The formalism is schematically described in Fig. \ref{fig_dmc_schematic}. 
The initial state for a trajectory is generated based on the thermal equilibrium polarization values of the electron (the nuclear spin polarization was assumed to be zero). \\
Due to the Markovian nature of the dynamics in the many spin system  only knowledge of the current state is sufficient to propagate the system in time. The Lindblad rates (Eqs. (\ref{eq_electron_effective_relaxation}) - (\ref{eq_dipolar_effective_interaction})) are first calculated based on the system parameters. The rate for the flip-flop interaction between an electron and a nuclei (Eq. (\ref{eq_hyperfine_effective_interaction})) is state-dependent and hence is initiated using the thermal equilibrium state. 
Next, two random numbers $r_1$ and $r_2$ are generated. The time required for the next event to take place is calculated using a natural logarithm of the first random number and the inverse of the sum of all rates of jumps that are possible in the system. In the next step it is decided which type of jump will be carried out. For this purpose all  rates are normalised to sum up to '1', and arranged in a cumulative sum array. An element in this array is identified which is higher in value than the second random number. The operation corresponding to that rate in the array element is carried out. Accordingly, operations with higher rate values are statistically more likely to occur than events with low rates. Once the operation or jump is carried out, the effective rates for the flip flop jump between a nucleus and the electron (Eq. (\ref{eq_hyperfine_effective_interaction})) are re-calculated based on the current system state. The steps outlined are repeated until a desired time is reached in the trajectory. Many trajectories are computed and averaged to approximate the polarization dynamics of the system. The method does not require vasts amounts of memory since only an $N$-element binary array is required that stores the current orientation of the $z$-component of each individual spin. The KMC method was implemented using MATLAB on a 3GHz 20 core desktop computer. \footnote{The full code can be made available by contacting the authors}  \\ \vspace{10pt} \\
A comparison of the polarization dynamics was carried out for the model spin system described previously with  $a = 5$ \AA\; inter-spin separation using the same parameter set as previously listed.
\begin{figure}[H]
\includegraphics[scale=0.2]{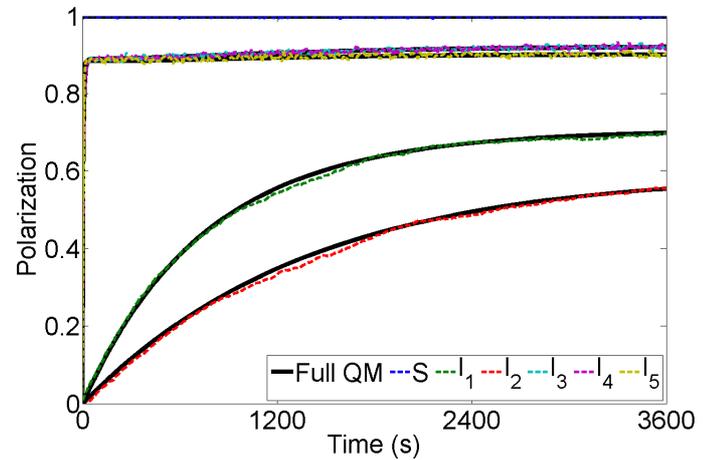}\caption{Comparison between Zeeman master equation and KMC}\label{fig_dmc_10k_Zeeman_polarization}
\end{figure}
\begin{figure}[H]
\includegraphics[scale=0.2]{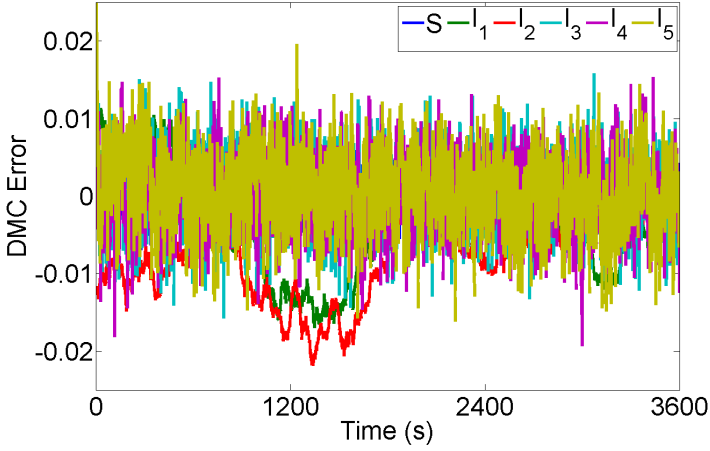}\caption{Error between Zeeman master equation and KMC}\label{fig_dmc_10k_Zeeman_error}
\end{figure}
Fig. \ref{fig_dmc_10k_Zeeman_polarization} shows polarization build-up curves computed using the Zeeman master equation (Eq. (\ref{eq_Zeeman_master_equation}), black solid lines) overlayed with the polarization build-up curves approximated by using the KMC algorithm (coloured, dashed lines). A total of 10 000 trajectories was averaged in the KMC procedure. Fig. \ref{fig_dmc_10k_Zeeman_error} shows the difference between the results of the two methods. The KMC polarization curves match the curves obtained by solving the Zeeman master equation very well, but are subject to a random error. This is an inherent property of the KMC algorithm and its dependence on random events. The difference does not generally exceed 2\% in polarization value, with the exception of early times. Averaging over a larger number of trajectories decreases the random error seen in the polarization curves obtained from the KMC method.

\section{Central electron model consisting of many nuclei}

To gain further insight into the optimal conditions for SE DNP for achieving the highest nuclear spin polarization and the fasted buildup rate simulations with many nuclear spins were carried out. This kind of analysis provides important information for a tailored radical design in which radical compounds are synthesized with desired properties. To analyze the spin dynamics during SE DNP in a central electron system containing many nuclear spins, a cubic grid of 124 $^{13}$C spins with a nearest neighbour separation of 4\AA \hspace{0pt} was chosen with 1\% uniform randomization in position of each nuclei. Such average distance corresponds to  26 M of a ${}^{13}$C labeled molecule or in other words it is slightly shorter then the average distance between ${}^{13}$C nuclei in free pyruvic acid (4.8 \AA) \cite{Ardenkjaer08}. The magnetic field was set to be 3.4T, the temperature was set 1K, and a microwave field amplitude of 20 kHz was chosen to approximate realistic experimental conditions. The relaxation parameters were: T$_1^{(S)}$ = 0.5 s, T$_2^{(S)}$ = 10 $\mu$s, T$_1^{(I)}$ =1 h, T$_2^{(I)}$ = 0.5 ms. The polarization dynamics for one hour of microwave irradiation was simulated using our KMC method.\\

\begin{figure}[H]
\includegraphics[scale=0.19]{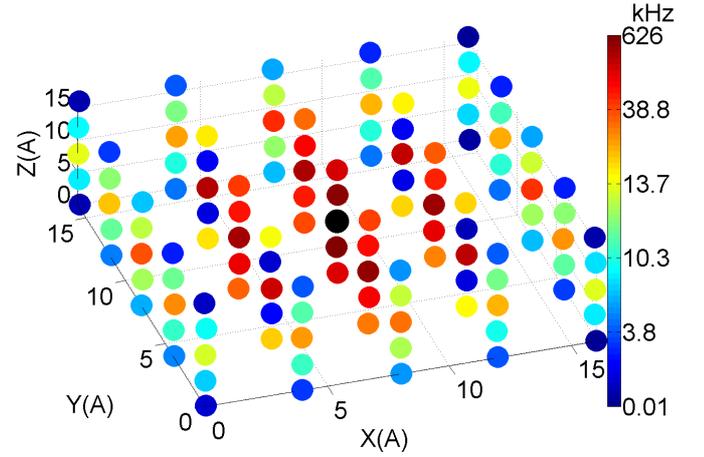}\caption{Strength of the secular hyperfine interaction $A_k$ between the central electron (black) and the 124 ${}^{13}$C nuclei. The colour scale indicates the coupling strength in kHz.}\label{fig_125spin_polarization2}
\end{figure}

\begin{figure}[H]
\includegraphics[scale=0.19]{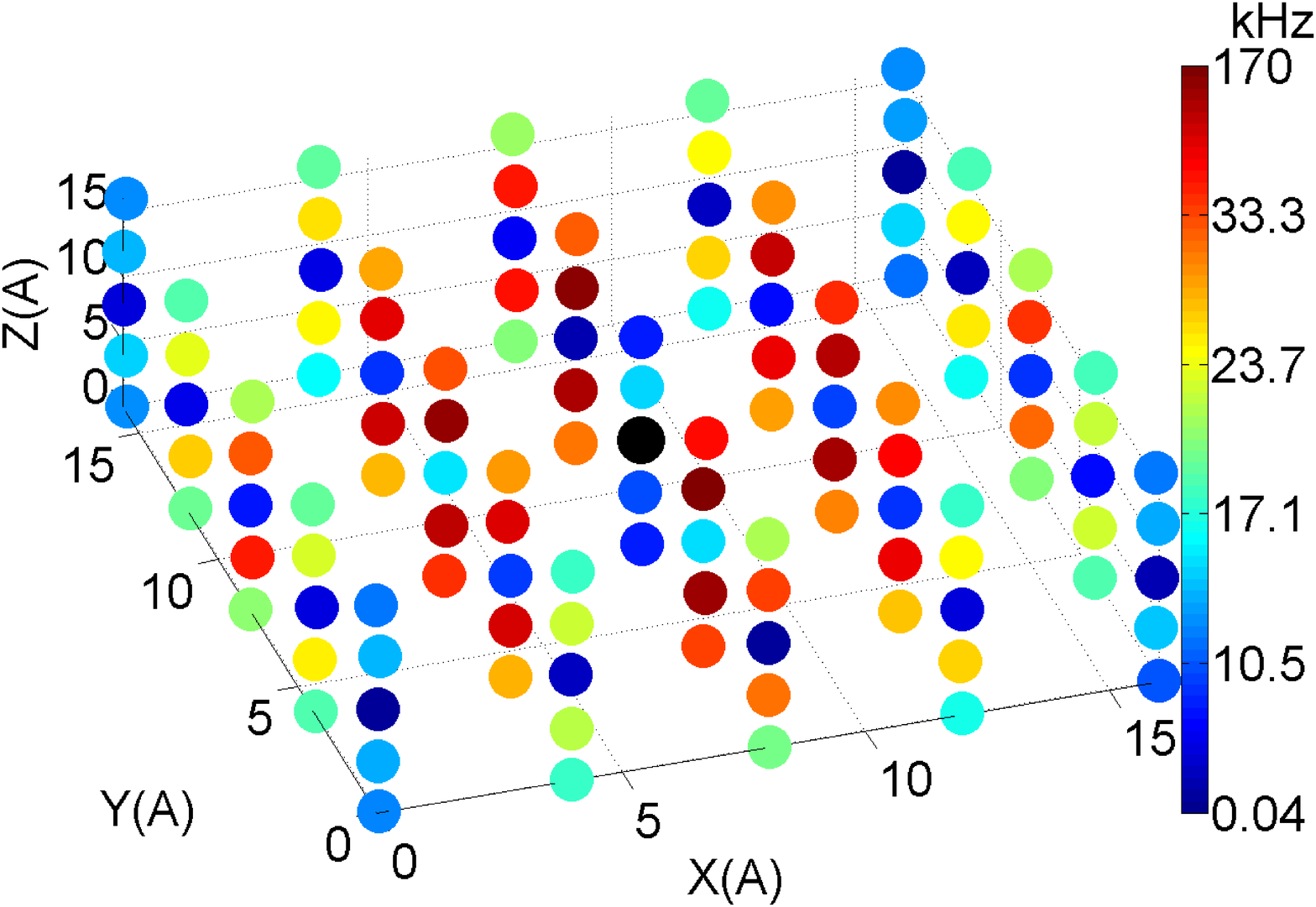}\caption{Strength of the pseudosecular hyperfine interaction $B_k$ between the central electron (black) and the 124 ${}^{13}$C nuclei. The colour scale indicates the coupling strength in kHz.}\label{fig_125spin_polarization3}
\end{figure}

Figs. \ref{fig_125spin_polarization2} and \ref{fig_125spin_polarization3} show the spatial dependence of the secular and pseudosecular hyperfine interaction strengths $A_k$ and $B_k$ for the 124 ${}^{13}$C nuclei.
The magnitudes of the secular part and the pseudosecular part of the hyperfine interaction are given by 

\begin{eqnarray*}
A_k &=& \left \vert \frac{\gamma_S \gamma_I}{r^3_k} \left ( 1 - 3 \cos^2 \theta_k \right ) \right \vert, \\
B_k &=& \left \vert - \frac{3}{2} \frac{\gamma_S \gamma_I}{r^3_k} \sin \theta_k \cos \theta_k \right \vert,
\end{eqnarray*}
where $\gamma_S$ and $ \gamma_I$ are the electron gyromagnetic constant and the nuclear gyromagnetic constant, respectively, $r_k$ is the distance between the nucleus $I_k$ and the electron $S$  and $\theta_k$ is the angle that the distance vector connecting the electron with the nuclei forms with the static magnetic field (conventionally chosen to be along the $z$-direction). 
The strength of the secular part of the hyperfine interaction $A_k$ is strongest for nuclei at positions with angle $\theta = 0$ or $\theta = \pi$. Conversely, the strength of the pseudosecular part of the hyperfine interaction is zero at these positions and strongest at positions characterized by $\theta = \pi/4$ or $\theta = 3\pi/4$. The interaction strength scales also with $r_k^{-3}$, so drops off relatively quickly with increasing distance between nuclei and the electron.

\begin{figure}[H]
\includegraphics[scale=0.19]{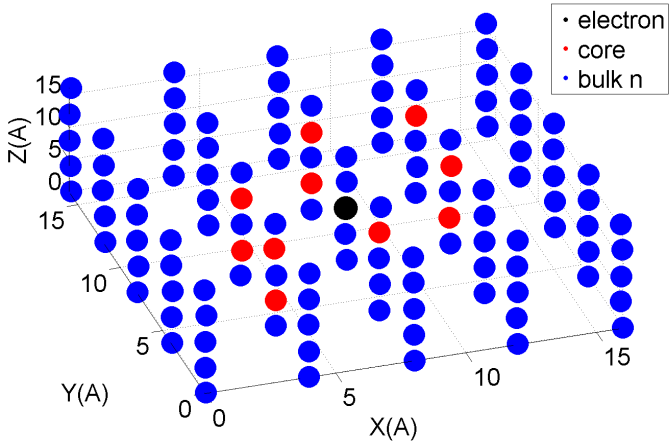}\caption{The nuclei in the ensemble around the central electron belong either to the core (red, effective rate for flip-flop jumps between electron and nucleus, Eq. (\ref{eq_hyperfine_effective_interaction}), is  higher than any of the effective inter-nuclei flip-flop rates, Eq. (\ref{eq_dipolar_effective_interaction})) or to the bulk (blue, effective rate Eq. (\ref{eq_hyperfine_effective_interaction}) is smaller than effective rate Eq. (\ref{eq_dipolar_effective_interaction})). Note that because of the 1\% uniform randomization of the position the configuration is not fully symmetric. }\label{fig_core_bulk}
\end{figure}

By calculating the effective rates, Eq. (\ref{eq_hyperfine_effective_interaction}), for the flip-flop jumps between the electron $S$ and the nucleus $I_k$ and comparing it to the effective rates with which the nucleus $I_k$ can carry out flip-flop jumps with adjacent nuclei, Eq. (\ref{eq_dipolar_effective_interaction}), it is possible to determine the nuclei that will predominately interact with the electron. We call these the core nuclei. All the remaining nuclei belong to the bulk (Fig. \ref{fig_core_bulk}).

\begin{figure*}
\includegraphics[width= \textwidth]{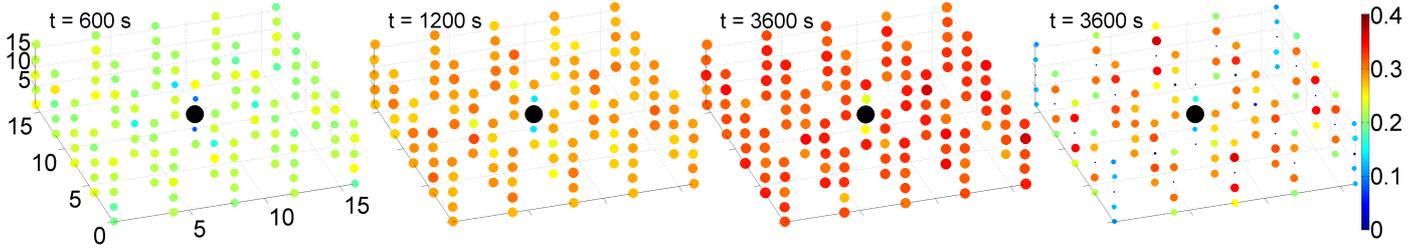}\caption{A) The distribution of the nuclear spin polarization at various time points during the buildup. A careful comparison with Fig. \ref{fig_125spin_polarization3} reveals that the nuclei with the strongest $B_k$ values have in the initial phase the highest polarization.  B) Nuclear steady-state polarization if the nuclear dipolar interaction coefficients $d_{ij}$ are set to zero . The nuclei are only polarized via the pseudo-secular part of the hyperfine interaction}\label{fig_composite_buildup}
\end{figure*}

Fig. \ref{fig_composite_buildup} shows the buildup of nuclear spin polarization in the system at various points in time. As expected the highest polarization increase is for nuclei with the highest strength of the pseudosecular interaction $B_k$. The polarization is also distributed by effective spin diffusion to regions in which $B_k$ is very small. This can be shown by setting the effective rate for internuclei flip-flops to zero  in the simulation. In this case only nuclei in regions with $B_k \ne 0$ receive significant electronic polarization during SE DNP.

\section{The effect of nuclei close to the electron}
\begin{figure}[H]
\includegraphics[scale=0.2]{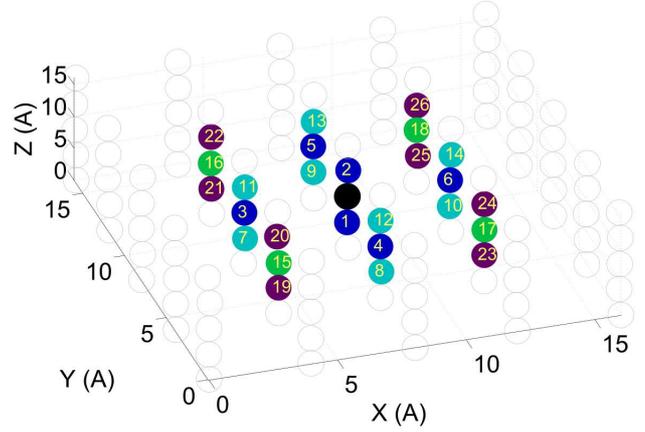}\caption{The order by which the layer of nuclei adjacent to  the electron was removed in successive simulations is indicated by the numbers. Different colour coding was used to group nuclei in four groups: The first six nuclei with strong secular hyperfine interaction and weak pseudosecular hyperfine interaction (dark blue), eight nuclei with relatively strong pseudosecular interaction (light blue). Four nuclei with strong secular hyperfine interaction and weak pseudosecular hyperfine interaction strength (green) and eight  nuclei with relatively strong pseudosecular interaction strength (purple) }\label{fig_125spin_polarization4}
\end{figure}

We analyze now the effect on the average bulk polarization if the configuration of this model system is changed  by first removing the six nuclei with the highest secular hyperfine interaction strength ($A_k$) and then subsequently also further nuclei adjacent to the electron. To the best of our knowledge a study as such is only possible with simulations involving large many-body systems, using the KMC algorithm and Zeeman-subspace master equation. The idea here is to understand the role of the secular and pseudosecular term of the hyperfine interaction of nuclei in immediate proximity of the electron for the bulk nuclear polarization. Every time one nucleus was removed the polarization dynamics were simulated starting from the initial thermal equilibrium state and the average nuclear spin polarization per nucleus has been calculated for the steady-state.  Fig. \ref{fig_125spin_polarization4} shows the order and the position of the removed nuclei.

\begin{figure}[H]
\includegraphics[scale=0.2]{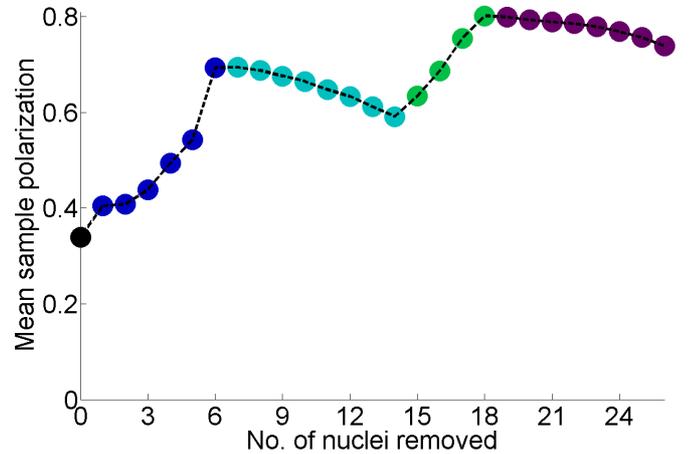}\caption{Mean polarization per nucleus versus the number of nuclei removed from the model system. The colour coding is identical to the one used in fig. \ref{fig_125spin_polarization4}. The first data point corresponds to the mean polarization without any nucleus removed (black). The next six data points (dark blue) correspond to the mean polarization if the six nuclei with strongest secular hyperfine interaction strength $A_k$ are removed. The next eight data points (light blue) corresponds to the mean polarization if nuclei with strong pseudosecular interaction strength are removed ($B_k$). The next four data points again corresponds to removal of nuclei with high $A_k$ values and the next eight data points (purple) corresponds to removal of nuclei with high $B_k$ values. The broken black line has been added as a visual guide only} \label{fig_125spin_polarization6}
\end{figure}

The dependence of the mean nuclear spin polarization per nucleus at steady state on the number of the removed nuclei is summarized in fig. \ref{fig_125spin_polarization6}. The first date point (black) represents the average nuclear spin polarization in the system without removal of any nuclei. The next six data points (dark blue) indicate the average nuclear spin polarization if the six nuclei with the strongest secular hyperfine interaction strength $A_k$ are successively removed. These are the nuclei just directly above and below the electron and next to the electron in the $xy$-plane in which the electron is located (see fig. \ref{fig_125spin_polarization4} for the position of the removed nuclei and fig. \ref{fig_125spin_polarization2} for the $A_k$ values). The pseudosecular interaction strength $B_k$  of these six nuclei are all very low (see fig. \ref{fig_125spin_polarization3}) and none of the removed nuclei belongs to the core nuclei group with high effective transfer rates of polarization from the electron (fig. \ref{fig_core_bulk}). The average nuclear spin polarization improves by more than a factor 2 due to the removal of these nuclei. The next eight nuclei that were successively removed (light blue colour) have the highest pseudosecular interaction strength $B_k$ and also relatively high $A_k$ values (see see figs. \ref{fig_125spin_polarization4}, \ref{fig_125spin_polarization2} and \ref{fig_125spin_polarization3}). Removal of these nuclei decreases the average nuclear spin polarization by about 15 \%.  Seven of these eight nuclei belong to the core nuclei group. The next four nuclei (green) that were removed have high $A_k$ values and low $B_k$ values. The removal of these nuclei increases again the average nuclear spin polarization in the ensemble. Compared to the polarization level of the last bright blue data point  the increase is more than 30\%. The polarization level compared to the average polarization for the reference system without any nuclei removed is increased by more than a factor of 2.3. The last eight nuclei that were removed have high $B_k$ values and the average nuclear spin polarization level decreases by about 10\%. \\

\begin{figure*}
\includegraphics[width = \textwidth]{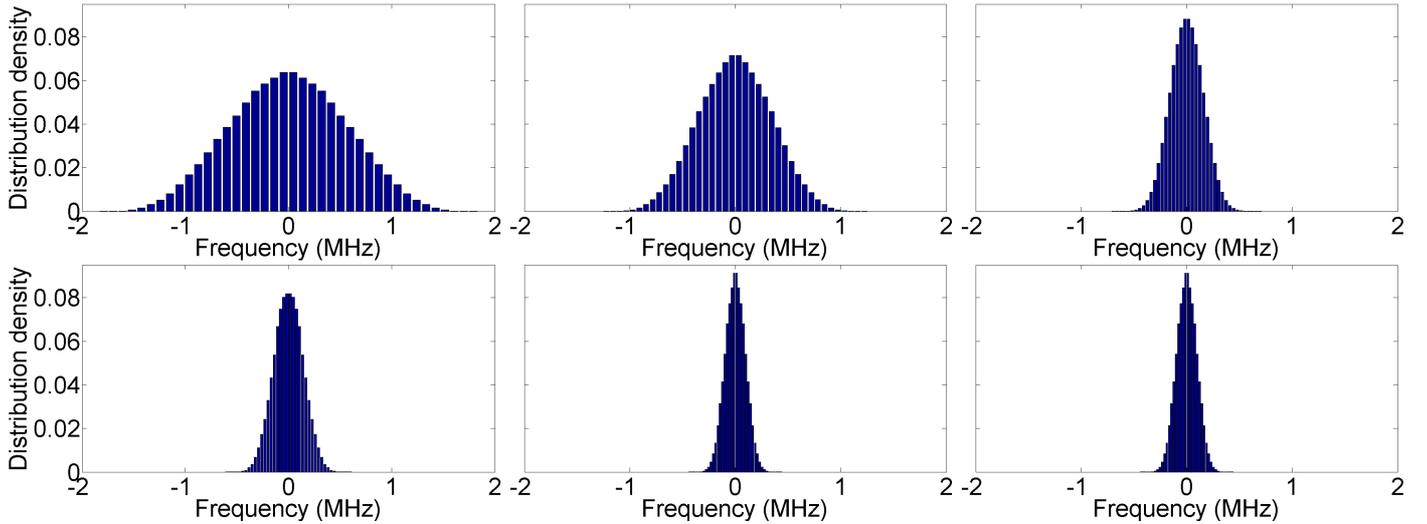}\caption{Histograms shows the distribution of frequencies for ZQ transitions that occur in the model spin ensemble. The frequencies are given as offset frequencies in respect to the ZQ transition $\omega_S + \omega_I$.  Successively nuclei are removed (none, 2, 6, 14, 18 and 24 removed) and the frequencies for ZQ transitions are calculated. First two of the nuclei with the strongest secular hyperfine interaction $A$ were removed followed by additional nuclei in descending order given by their secular hyperfine interaction strength. Fig \ref{fig_125spin_polarization4} shows the positions of the removed nuclei. Note that the width of the distribution becomes initially narrower with increasing numbers of nuclei removed. However, since the six nuclei that were removed at the end have only very weak $A_k$ values there is no significant change between the last two histograms}\label{fig_125spin_polarization5}
\end{figure*}

In order to  interpret these results it is instructive to analyze the distribution of frequencies which are characteristic for SE DNP. The effect of the nuclei with  strong secular hyperfine interaction (high $A_k$ values)  is a splitting of the frequencies $\omega_S \pm \omega_I$ at which SE DNP is mediated by an excitation of ZQ or DQ transitions \cite{Nagarajan10, Vega1}. The splitting of these frequencies for each nuclei $I_k$ can be calculated by using \cite{Nagarajan10, Vega1}
\begin{equation}
\omega^{DNP}_{I_k} = \omega_S  \pm \omega_I - \frac{1}{2} \sum_{i \ne k} s_i  A_i, \qquad s_i = \pm 1,  
\label{dnpfreq}
\end{equation}
where $s_i$ is the sum means that one needs to add up over all possible permutations. Because of the very high number of frequencies that will result from the coupling of 124 nuclei to the electron we have selected only 25 nuclei to calculate these frequencies. The 25 nuclei were identified by ordering the nuclei with decreasing secular hyperfine interaction strength $A_k$ and by selecting the first 25 with relatively high $A_k$ values. The distributions of frequencies calculated using Eq. (\ref{dnpfreq}) are shown in histograms (see fig. \ref{fig_125spin_polarization5}). In these histograms the fractional number of transition frequency at which nucleus-electron flip-flop occur are plotted against the frequency offset to the ZQ frequency $\omega_S + \omega_I$.  Note that with removal of the first six nuclei the distribution becomes significantly narrower. Any further removal of nuclei reduces the width of the distribution further until the $A_k$ values of the removed nuclei are so weak that no further change can be observed. From fig. \ref{fig_125spin_polarization5} we can conclude that initially successive removal of nuclei increasingly causes nuclei to have ZQ  transition frequencies closer to the microwave carrier frequency set to $\omega_S + \omega_I$ and hence SE DNP will become more efficient. However, while this has a positive effect on the average nuclear spin polarization, removal of nuclei with high $B_k$ values that can pass polarization on to the bulk by spin diffusion has a negative effect of the average nuclear spin polarization. These considerations show the delicate interplay between the various parameters in the spin ensemble.  

We observed in our model simulations that the average nuclear spin polarization is more than 2 times higher if the 26 nuclei adjacent to the central electron are removed. The removal of the nuclei means that nearest distance between nuclei and electron has increased  by a factor of 2. If further layers of nuclei are removed it is to be expected that the average nuclear spin polarization will decrease to levels below the inital level of the system without any nuclei removed since the pseudosecular interaction strength becomes too weak for efficient polarization transfer between the central electron and the interacting nuclei. From these considerations it is clear that there must be a set of parameters that gives the optimum DNP efficiency and for which the nuclear spin polarization will reach its highest level. Such an optimum does depend in a complex way on all parameters including the microwave irradiation strength, the average distance between the nuclei,  the nearest distance between nuclei and the central electron and the various relaxation parameters. \\

\section{Extension to systems with more electrons}
With the success of the Zeeman projection for SE DNP dynamics, we are currently pursuing a similar approach for the cross effect DNP  mechanism (CE DNP), where electron pairs play a key role, leading to polarization dynamics more efficient than in the SE DNP case \cite{Kessenikh64, Hwang67}. Using again two adiabatic eliminations it is possible to first project the dynamics onto a subspace that contains all operators that commute with the nuclear Zeeman operators $I_{kz}$ and the Zeeman operator of one of the electrons \cite{us2}, and subsequently project it on to the Zeeman subspace to obtain a master equation that can be written in the Lindblad form. However, in contrast to the Lindblad master equation for SE DNP, Eq. \ref{eq_Zeeman_master_equation}), the set of jump operators contains additional three spin operators that describe flip flops between two electrons and a concomitant change of orientation of the nuclear spin. The Lindblad master equation for CE DNP can also be used in a wide parameter space but stricter conditions apply in comparison to SE DNP. Further extensions of this formalism to the case of rotating solids for simulations of DNP during magic angle spinning are also feasible. In addition, it was shown recently that also the dynamics of a system containing several electrons and one nuclear spin can be described by a master equation in Lindblad form \cite{Luca2015}.\\
There are only few experiments published in the literature where SE DNP seems to be the sole mechanism for generating the non-thermal nuclear spin state. Seminal work by Vega and coworkers has demonstrated that rather a mixture of SE DNP and CE DNP is responsible for the build up of nuclear polarization at cryo-temperatures using the trityl radical \cite{Hovav15, Banerjee13}. Even for such complex systems it appears to be feasible, based on our work involving the reduction of the required state space, to derive a set of equations that can describe the nuclear polarization dynamics.

\section{Radical design}
The simulations presented in this work point to a possible explanation why the trityl radical currently frequently used for ${}^{13}$C dissolution DNP seems to perform particularly well \cite{Ardenkjaer98, Johanneson09}. The explanation is based on the assumption that there is also an optimal distance for CE DNP between the electron and the nuclei. The bulky aromatic groups of the tritly compound keep the molecules that carry the ${}^{13}$C-label (e.g. [$1-{}^{13}$C] pyruvate) at an appreciable distance to the electron (for an analysis using ${}^1$H see \cite{Trukhan13}).  The relatively high value of the nearest distance between ${}^{13}$C nuclei and the electron results in a relatively narrow distribution of the possible DNP transition frequencies in the nuclear spin ensemble but on the other hand the ${}^{13}$C nuclei are still near enough to ensure a high pseudosecular interaction between the first layer of nuclei and the electron. Furthermore, in this case the difference between the secular hyperfine interaction strength of adjacent nuclei will not be as high as for a configuration in which the nuclei can get very close to the electron. This in turn will lead to more efficient spin diffusion (rate Eq. (\ref{eq_dipolar_effective_interaction})) between nuclei closer to the electron and the next layer further away. These considerations demonstrate that the performance of DNP is critically dependent on the immediate molecular environment of the paramagnetic centre and that improvements in the performance of DNP could be obtained by specific radical design.

\section{Conclusion}
We have demonstrated the validity of a formalism, which allows simulations of large many-spin systems. We tested the accuracy of the approach for a small spin system with a varied inter-spin separation. Our results show that the Zeeman projection agrees with the full quantum-mechanical master equation down to 2 \AA \; separations, where the maximal observed relative error was less than 3\%. The projection is valid even for parameters which might be considered as extreme in cases of DNP experiments. We show also that simulations relying on the Kinetic Monte Carlo algorithm are in agreement with the Zeeman-projected master equation, subject to random error, which is reduced for simulations with higher numbers of trajectories. The classical KMC algorithm presents a powerful tool, suitable for use in any system where the dynamics are incoherent and resemble classical behaviour.\\
We show the necessity and advantage of using large, many-spin simulations when studying the mechanism of DNP by analyzing the effect of the nuclei in immediate vicinity to the electron on the DNP performance. We studied the effect of removing core-nuclei from a virtual cubic lattice system on polarization build-up rates of remaining nuclei, as well as their steady-state values. A clear improvement in both can be seen when core nuclei - those in the immediate vicinity of the electron, are removed from the system. Build-up rates increase significantly, and the mean steady-state polarization level increases by more than a factor of 2. We explain this using simulated DNP transition spectra, which show linewidth narrowing with the removal of nuclei strongly coupled to the electron. Crucially, the improvement can be explained using the effective rates previously derived. The results presented here may give an insight or guidance in radical-design for optimised DNP experiments. The efficiency of designed radicals can be theoretically predicted and assessed.\\
\section{Acknowledgements}
This work was funded by the UK Engineering and Physical Science Research Council by a grant to WK (EP/I027254)



\bibliographystyle{elsarticle-num} 
\bibliography{JMR_submission}





\end{document}